**Global temperature goals should determine the time horizons for greenhouse gas emission metrics**

S. Abernethy[1,2,*], R. B. Jackson[2,3]

[1] Department of Applied Physics, Stanford University

[2] Department of Earth System Science, Stanford University

[3] Woods Institute for the Environment and Precourt Institute for Energy, Stanford University, Stanford, 94305, USA

* Corresponding author: sabernet@stanford.edu

Abstract

Emission metrics, a crucial tool in setting effective equivalences between greenhouse gases, currently require a subjective, arbitrary choice of time horizon. Here, we propose a novel framework that uses a specific temperature goal to calculate the time horizon that aligns with scenarios achieving that temperature goal. We analyze the Intergovernmental Panel on Climate Change Special Report on Global Warming of 1.5 °C Scenario Database to find that time horizons that align with the 1.5 and 2 °C global warming goals of the Paris Agreement are 24 [90% prediction interval: 7, 41] and 58 [90% PI: 41, 74] years respectively. We then use these time horizons to quantify time-dependent emission metrics with methane as our main example. We find that the Global Warming Potential values that align with the 1.5 and 2 °C goals are $GWP_{1.5\,°C} = 75$ [90% PI: 54, 107] and $GWP_{2\,°C} = 42$ [90% PI: 35, 54]; for the Global Temperature change Potential they are $GTP_{1.5\,°C} = 41$ [90% PI: 16, 102] and $GTP_{2\,°C} = 9$ [90% PI: 7, 16]. The most commonly used time horizon, 100 years, underestimates methane emission metrics by 34-38% relative to the values we calculate that align with the 2 °C goal and 63-87% relative to the 1.5 °C goal. To best align emission metrics with the 1.5 °C goal of the Paris Agreement, we recommend a 24-year time horizon, using 2045 as the endpoint time, with its associated $\text{GWP}_{1.5\,°C} = 75$ and $\text{GTP}_{1.5\,°C} = 41$.

**Introduction**

Emission metrics, the "exchange rates" between greenhouse gases, are necessary for policy-makers and decision-makers to compare the relative effects of different gases to carbon dioxide, the generally accepted baseline[1–10]. Emission metrics underpin the Nationally Determined Contributions (NDCs) of the Paris Agreement and form the basis of so-called "carbon dioxide equivalent" emissions by allowing greenhouse gases to be compared on the same scale. Here, we develop a framework to quantify emission metrics that align with the Paris Agreement goal to keep global warming well below 2 °C and ideally to 1.5 °C, compared to preindustrial levels. Currently, the Intergovernmental Panel on Climate Change (IPCC) uses 20- and 100-year time horizons for emission metrics[1] while simultaneously acknowledging that these arbitrary values "should not be considered as having any special significance"[11]. Our framework derives time horizons of special significance that align with specific temperature goals.

Time horizons are required for most emission metrics because greenhouse gases with shorter atmospheric lifetimes impact the climate much more over the short-term compared to the more constant effects of carbon dioxide over centuries[12]. We focus on methane because it is the second-most important greenhouse gas after carbon dioxide and its atmospheric lifetime of only 11.8 years means that its emission metrics depend strongly on the time horizon[1,13]. The extension of our method to other short-lived climate forcers would be straightforward.

The IPCC's use of two reference time horizons, 20 and 100 years, has been criticized for being arbitrary and unjustified[14], with the choice between the two often based on political perspectives or vested interests[15]. Quantitative justifications for specific time horizons are surprisingly lacking in the literature. One rare example comes from Sarofim and Giordano, who argued that time horizons could be justified based on their equivalent economic discount rate, and showed that the 100-year time horizon aligned implicitly with a 3.3% discount rate (consistent with many climate and economic impact analyses)[15]. Here, we instead define "justified time horizons" as those that align with the predicted timing of peak warming for the specific temperature goals of the Paris Agreement. We then use these justified time horizons to quantify time-dependent emission metrics.

What is the justification for aligning emission metrics with temperature goals? We believe that greenhouse gases should be weighted based on their contribution to achieving a given climate goal. We acknowledge that emission metrics are used in a wide range of applications, not all of which relate to temperature stabilization goals, meaning that the criteria for selecting time horizons is not universal. Potential policy goals, beyond temperature stabilization, include intergenerational equity[8,9,16], air quality[17], and cost-effectiveness[18,19]. However, since the main goal of the Paris Agreement is to limit the temperature peak well below 2 °C, the impact of emissions on achieving this specific temperature goal should dictate the relative importance of different greenhouse gases. We use 1.5 °C and 2 °C as representative examples of the temperature goals of the Paris Agreement since they are the two values explicitly mentioned. Throughout this work we use "justified" as short-hand for "justified in its alignment with the Paris Agreement temperature goals," while noting that other justifications could and should be developed for different applications.

Our framework is primarily designed for times before peak temperature; we make no claim about optimal emission metrics after the time of peak temperature, but simply aim to align emission metrics with the current goal of avoiding the exceedance of a given temperature limit. We hypothesize that as a temperature peak is approached, new climate goals will be developed (likely regarding stabilization at a certain temperature) and argue that future time horizons should be calculated that align with those goals.

We recognize the value in having an agreed upon emission metric to standardize the weighting of different greenhouse gases and support the narrowing of the broad range of possible emission metrics. The use of the Global Warming Potential with a 100-year time horizon (GWP100) as the common metric in the implementation of the Paris Agreement, as agreed upon in the 24th Conference of the Parties (COP24) in 2019, is a step in the right direction towards standardization[19]. However, it has been shown that pathways with identical carbon dioxide equivalent emissions calculated using GWP100 can vary in temperature by as much as 0.17 °C[20]. One proposed solution is simply to report both GWP20 and GWP100 in all cases, as is often done for city-highway fuel economy, for example. However, this practice still leaves the choice between the two as a subjective, arbitrary decision[21]. Another approach involves bounding the range for emission metrics using a "do no harm" principle[10], but this leads to a large range of potential values. Given the openness of the IPCC to revising emission metrics in future Assessment Reports, we believe our framework for selecting time horizons could help create more justified, but still standardized, emission metrics.

Generalizing, temperature changes near the peak temperature are more important than those that happen long before or after, in part because they introduce the possibility of passing climate tipping points[22]. With justified time horizons, as we define them, greenhouse gases would be weighted by their impact on achieving specific temperature goals at the time at which temperature peaks. To illustrate our point, consider an extreme hypothetical temperature goal to limit global warming to 4 °C (a threshold unlikely to be met for several decades, if ever). If a 10-year time horizon was used, this would drastically overvalue current emissions of short-lived climate forcers like methane and suboptimally distribute resources towards reducing them, thereby lowering the likelihood of achieving the temperature goal. Higher temperature goals are associated with longer time horizons for achieving them, and emission metrics should reflect this fact. Generally speaking, the more that rapid action towards stringent temperature goals (such as 1.5 °C) proves feasible, the higher weight we should give to short-lived climate forcers such as methane.

Our framework builds on the work of Manne and Richels[18], who established the time-dependence of emission metrics based on price ratios and claimed that the optimal emission metrics for methane depended on both the time and the temperature goal. We also incorporate the "combined target and metric approach" of Shine et al.[23] and Berntsen et al.[24], who clearly outlined the importance of aligning temperature goals with emission metric time horizons. They established that lower target temperatures imply an earlier target year—and emission metrics should reflect this fact. However, their work was limited because (in the words of Berntsen et al.) "a unique relationship between the target level and the target year [had] not been derived in the existing literature."[24] Using the current availability of scenarios from numerous Integrated

Assessment Models (IAMs)[25], we derive the first relationship of this type. The works closest to ours are those of Tanaka et al.[19,26,27]. They first examined time-dependent emission metrics using one IAM and three scenarios for stabilization at 2, 3, and 4 °C[26]. They then presented a framework for choosing time horizons using cost-effective socioeconomic metrics, this time with five scenarios but again only one IAM[19]. We take a different approach by: 1) considering climate only, rather than socioeconomics, 2) aggregating ten IAMs instead of a single model for robustness, and 3) analyzing 213 mitigation pathway scenarios instead of five to generalize the framework for use with a general temperature goal and associated time horizon.

We emphasize the importance of increasing NDCs such that temperature stabilization can be achieved; all greenhouse gases must be reduced rapidly. Climate policy using our framework would first set limits on carbon dioxide-equivalent emissions to ensure that the long-term temperature goal is achievable[23]. Indeed, the emission metrics we present are only valid if a temperature peak occurs. Linking the time horizon with the temperature goal means that increased emphasis on methane would not reduce the importance of carbon dioxide, since it would simply make the overall climate target stricter[24]. The emission metrics we present, along with the future costs of different greenhouse gas mitigation and removal approaches, would then dictate the portfolio of emission reductions needed to most efficiently achieve the specified temperature goal.

**Emission metrics**

Depending on the choice of emission metric, the implied severity of methane relative to carbon dioxide ranges from 4 to 200, although most estimates are between 10 and 80[28]. Our framework narrows this range by incorporating temperature goals and the temporal proximity to the peak temperatures that meet those goals.

The most commonly used metric for comparisons between greenhouse gases, the Global Warming Potential (GWP), measures the radiative forcing of a pulse of gas relative to the radiative forcing of a pulse of carbon dioxide with equal mass, both integrated over a given time horizon. However, GWP has been criticized widely, including that it is inappropriate for the temperature goals of the Paris Agreement because it quantifies radiative forcing rather than temperature[18,23,29–33].

The Global Temperature change Potential (GTP) is a more policy-relevant metric that quantifies the temperature impact of a pulse of gas relative to the temperature impact of a pulse of carbon dioxide with equal mass, both evaluated at an endpoint time horizon[23,32]. Limiting the instantaneous peak temperature is the explicit goal of the Paris Agreement due in part to the significance of instantaneous elevated temperature in causing heat waves and extreme events[34]. However, the policy relevance of GTP comes at the expense of increased uncertainty because climate modelling is required to calculate it, unlike the purely physical GWP[23,35,36].

Extended periods of time at elevated temperature directly impact humanity and the climate, including through sea-level rise and the potential passing of tipping points in the climate system such as melting sea-ice and permafrost thawing. The integrated Global Temperature change Potential (iGTP) quantifies the duration of elevated temperature by integrating the temperature impact over time (and as such is based on the same climate modelling as GTP)[37,38].

A growing body of research suggests that pulse emissions of carbon dioxide should not be equated with pulse emissions of short-lived climate forcers like methane, but rather with sustained changes ('steps') in their emission rates[1,23,39]. This approach, using 'step-pulse' emission metrics like CombinedGTP[40], captures the contrasting nature of carbon dioxide as a stock pollutant and methane as a flow pollutant and also reduces the sensitivity of emission metrics to the time horizon—but selecting a time horizon is still required. Another recent emission metric is GWP*, which equates pulses of carbon dioxide with emission rates of methane over a preceding time range (often 20 years)[10,29,30,33,41]. However, GWP* has been criticized for its unfairness on equity grounds at the country level[9,16] and for its potential incompatibility with the Paris Agreement due to the requirement to take preceding decades into account[19]. As is currently done by the IPCC, we focus on metrics that equate pulses with pulses, but note that our framework is also compatible with CombinedGTP (see Methods for details and an example).

Keeping global warming well below 2 °C is the main goal of the Paris Agreement, but other considerations beyond temperature are important and should be incorporated into decision-making processes. The social cost of methane, for example, is much larger than temperature-based emission metrics alone. This is because increased methane concentrations cause increased surface ozone which negatively impacts human health and crop productivity, among other secondary effects[12,17]. Two other emission metrics, the Global Damage Potential[42] and the Global Cost Potential[43], come from a cost-benefit and a cost-effectiveness framing of climate change, respectively. Economic metrics such as these are further along the climate change cause-effect chain, meaning that they are more descriptive of the actual damages caused, but at the expense of higher uncertainty due to the required modelling of the socioeconomic impacts of climate change[28]. Since our work focuses on temperature goals, we limit our focus to physical metrics (radiative forcing and temperature) rather than socioeconomic metrics. Therefore, we present GWP and GTP as representative emission metrics to outline our framework.

**Results**

We start by screening the 21st century mitigation pathway scenarios in the IPCC Special Report on Global Warming of 1.5 °C (SR1.5) Scenario Database[44] to extract the 213 scenarios with a temperature peak. We then aggregate all 213 scenarios that come from ten IAMs (Figure 1) to form the basis of our dataset. All scenarios are weighted equally. Next, we plot the peak temperature ($T_{peak}$) against the time at which it occurs ($t_{peak}$) to derive a relationship between these variables given by

$$t_{peak} = a\, T_{peak} + b, \tag{1}$$

where $a = 68 \pm 4$ years/°C and $b = 1943 \pm 7$ years (Figure 2). These parameters are calculated by performing linear least squares on the raw temperature peak data (Figures 1 and 2).

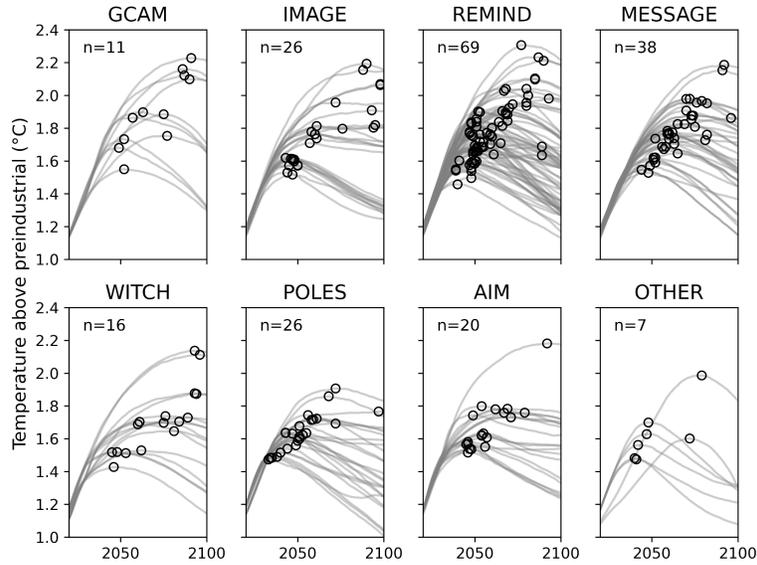

Figure 1: All 213 scenarios with a temperature peak from the IPCC SR1.5 Scenario Database[44], categorized by IAM used. The number of scenarios for each IAM (n) is shown in each panel. Temperature peaks are identified with black circles.

Equation 1, on its own, is simply a description of the emergent linear relationship between the peak temperature and the time at which it occurs (Figure 2). To understand the applicability of this relationship, we use the "very likely" nomenclature adopted by the IPCC to calculate 90% prediction intervals (PI), the range in which we predict 90% of future scenarios will fall. The 90% PI for the relationship between the temperature peak and its time of occurrence has a range of approximately 33 years for a given temperature goal (Figure 2).

Our estimates for when the temperature goals of the Paris Agreement will be reached are 2045 [90% PI: 2028, 2062] for 1.5 °C and 2079 [90% PI: 2062, 2095] for 2 °C. These results correspond to time horizons of 24 and 58 years, respectively, from present day (2021). The 100-year time horizon most commonly used by the IPCC lies far beyond these 24- and 58-year time horizons that align with their specific temperature goals. Our estimate for the time of reaching a 2 °C peak temperature, 2079, is 15 years beyond Tanaka et al.'s estimate of 2064[26], thereby revising their emission metrics for methane downwards by roughly 20% for GWP and iGTP (integrated metrics) and by 40% for GTP (an endpoint metric).

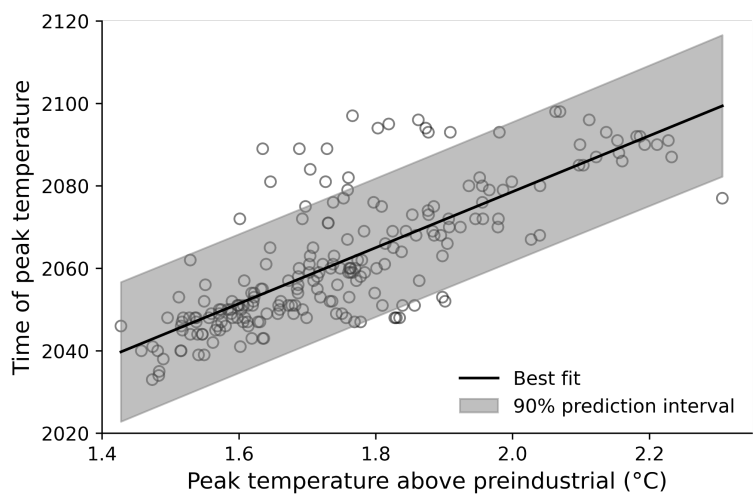

Figure 2: The relationship between peak temperature and the time the peak temperature is reached for all 213 scenarios and all 10 IAMs. Individual scenarios are shown as points, as in Figure 1. Overlaid on this raw data is the line of best fit (black line) and 90% prediction interval (gray shaded area).

We then calculate time-dependent emission metrics using the established "combined metric and target approach"[24] where the time horizon is the proximity to the time of peak temperature (Figure 3a)[26]. Methane's emission metrics increase with proximity to the time of peak temperature, meaning that the values we present in this paper are specific to 2021 (and will increase in the future). Closer to the time of peak temperature, climate goals may change. For example, a goal to ensure long-term stability at a certain temperature would lead to emphasizing carbon dioxide, whereas a goal to rapidly return to preindustrial temperatures would emphasize short-lived climate forcers. We focus solely on times before the peak temperature where the explicit end goal (a temperature limit in the Paris Agreement) has been formalized; times beyond the peak are addressed in the Discussion but are generally beyond the scope of this paper.

Combining these established time-metric relationships (Equations 2-4 and Figure 3a) with our novel temperature-time relationship (Equation 1 and Figure 2), we derive a primary result: a relationship between temperature goals and emission metrics (Figure 3b). We then use our framework to quantify emission metrics for the specific temperature goals of the Paris Agreement (Table 1). We propose the use of subscripting the temperature goal for an emission metric. For example, $GWP_{2\,°C}$ is the value for GWP that aligns with scenarios satisfying a 2 °C temperature goal and, for methane, is equal to 42 [90% PI: 35, 54].

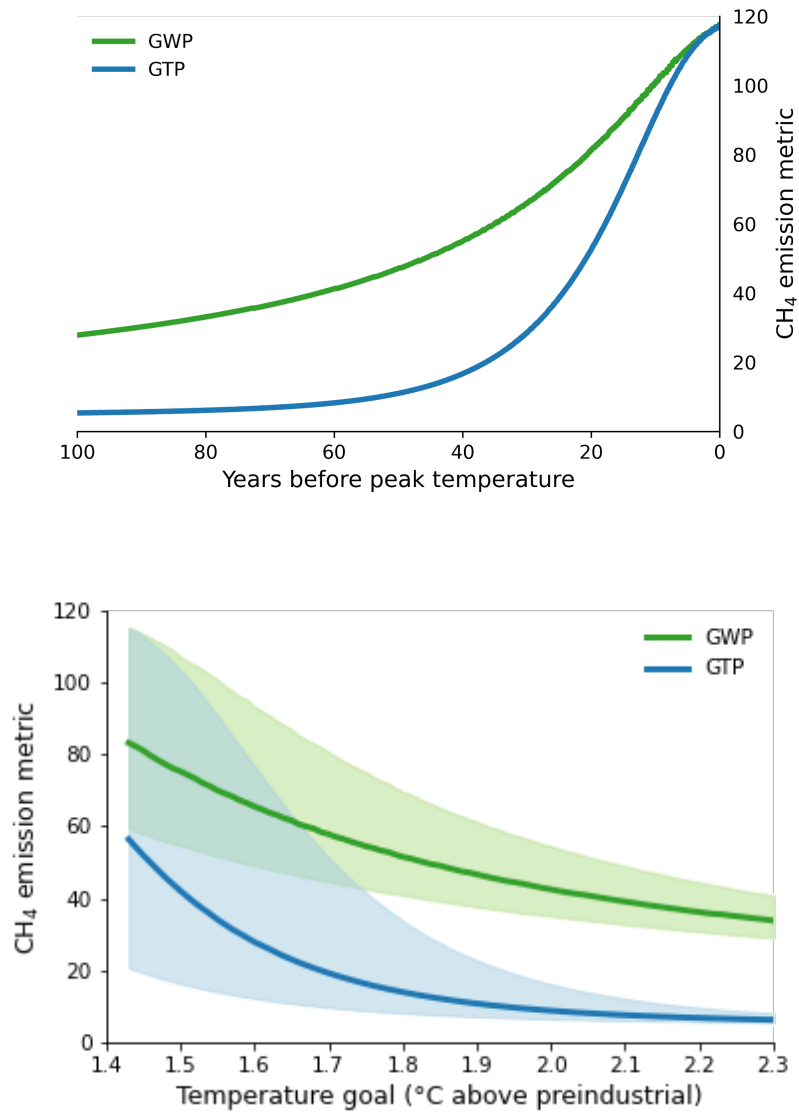

Figure 3: Time dependent emission metrics (GWP in green and GTP in blue) for methane. Panel a) The "combined target and metric approach"[24] where the time horizon is the number of years before peak temperature. Panel b) Combining this approach with the results of Figure 2 to show a relationship between temperature goal and emission metrics. Shaded area is the 90% prediction interval. See Methods for details and analytic equations[1,45].

The time horizon we calculate to align with scenarios peaking at 1.5 °C, 24 years, is close to the commonly used 20-year time horizon. Using our framework, therefore, the use of a 20-year time horizon can roughly be thought of as implicitly aligning with the 1.5 °C temperature goal. On the other hand, the commonly used 100-year time horizon is far beyond the time horizon we calculate that aligns with the 2 °C goal (58 years). In fact, if Equation 1 (the linear relationship shown in Figure 2) could be extrapolated to higher temperature peaks, a 100-year time horizon would roughly align with a 2.6 °C temperature stabilization. (We address the question of whether a linear relationship is the best choice for fitting and extrapolating this data in the Discussion.) When the 100-year time horizon is applied to a 2 °C goal, instead of our proposed 58-year time horizon, methane emission metrics are significantly underestimated relative to those we find that align with the 2 °C goal: 28 vs. 42 for GWP (34% lower) and 5.4 vs. 8.7 for GTP (38% lower). The 100-year time horizon underestimates emission metrics that align with the 1.5 °C goal even more drastically: by 63% for GWP and 87% for GTP.

Table 1: Summary of the time horizons and emission metrics for specific temperature goals of the Paris Agreement. 90% prediction intervals are shown in brackets.

|  | **1.5 °C** | **2 °C** |
|---|---|---|
| **Time horizon (years)** | 24 [7, 41] | 58 [41, 74] |
| **GWP** | 75 [54, 107] | 42 [35, 54] |
| **GTP** | 41 [16, 102] | 9 [7, 16] |
| **iGTP** | 78 [56, 109] | 43 [36, 56] |
| **CGTP (years)** | 1850 [552, 2436] | 2738 [2436, 2929] |

**Discussion**

We present recommended time horizons for emission metrics but cannot address which emission metric should be used in all cases. This choice, or multiple choices based on different goals, must be made based on the outcome of interest, the decision between endpoint and integrated, and the tradeoff between the usefulness of temperature-based or socioeconomic emission metrics with the added uncertainty that comes with the modelling required to calculate them[6,28]. We believe (i)GTP is more applicable than GWP when considering temperature goals, but recognize the potential for our framework to be used for other goals. For radiative forcing targets, GWP would be better suited; if policy decisions between pulse emissions of carbon dioxide and sustained changes in methane emission rates were being considered, a step-pulse metric such as CombinedGTP[40] might be the metric of choice. Aiming for net zero emissions is also an explicit goal of the Paris Agreement, but it has been shown that the definition of net zero emissions depends heavily on the metric used[31,46]. Regardless of the chosen metric, however, our framework avoids the subjective, arbitrary choice in time horizon (often either 20 or 100 years) by determining the time horizon that aligns with a given climate goal.

The relationship between peak temperature and the time at which it occurs (Figure 2) is to our knowledge the first of its kind but comes with several limitations. First, the assumption of linearity in the temperature-time relationship (Equation 1) is not motivated by an underlying physical mechanism. The outliers that fall outside of the prediction interval (roughly between 1.6 and 1.9 °C) are primarily above this interval, suggesting that a fit with a negative second derivative may better capture this relationship. However, fitting the data in Figure 2 with a quadratic or logarithmic function changes the predicted time of peak warming by less than 2 years for all temperature goals relevant to the Paris Agreement (1.5-2 °C). Therefore, we make no claim about the applicability of this relationship to temperature goals extrapolated beyond these temperature goals, but note that the functional form of the fitted relationship has only a minor influence on the predicted timing of peak temperature within this range. Second, the set of scenarios in SR1.5 is not exhaustive (and indeed there are arguments calling for IAMs to be supplemented with other modelling approaches[47]). SR1.5 does not include scenarios with temperature peaks higher than 2.3 °C, thereby missing a broad class of likely scenarios (based on current global NDCs) that may reach 3 °C warming[48]. The SR1.5 scenarios are also primarily 'peak and decline' scenarios, whereas 'stabilization' scenarios may follow a qualitatively different temperature trajectory[46]. Finally, the SR1.5 scenarios end in 2100, thereby potentially biasing the prediction interval downwards (earlier). To address these concerns and increase the robustness of our results, we encourage the extension of the scenario database to include higher temperature peaks, scenarios that continue past 2100, and more extensive modelling of the full space of potential future scenarios (ideally weighted based on their likelihood).

One potential criticism of our framework (and other work upon which it builds[18,23,24,46]) is that it is valid only until the time of peak temperature. Furthermore, it implicitly ignores temperature impacts that occur after the peak. This criticism is valid, but we argue that it is best aimed toward the creation and modification of climate goals, rather than at the specifics of our framework. From a cost-effectiveness perspective, it has been shown that emission metrics stabilize asymptotically as temperature stabilizes[19], meaning that the resolution of this criticism may be the simplest solution: to take our recommended emission metric at the time of peak for all times thereafter. One could also develop a framework similar to the one we present where the goal is to minimize time at elevated temperatures above 2 °C, for example, or to minimize the rate of temperature increase to allow ecosystems more time to adapt[34]. Whether to use the time horizons we present that align with the temperature peak goals, or to have different emission metrics before and after the temperature peak, is a policy-specific question beyond the scope of this work. We welcome future extensions of our framework to align with alternative climate goals, particularly those that incorporate times after peak temperature.

The application of our results in climate policy is complicated by the fact that we consider time-dependent emission metrics (where, for methane, they strictly increase as the time horizon decreases). While such a dynamic approach may have inhibited the early adoption of time-dependent emission metrics[18], and may prove politically difficult to implement, we believe that dynamic carbon pricing already signals an ability for government agencies to use time-dependent emission metrics for policy applications. More research is needed into the political feasibility of updating emission metrics over time, such as with the global stocktake every five years from 2023 onwards.

Although NDCs and long-term national pledges are currently insufficient to keep warming below 2 °C, let alone 1.5 °C[48–50], the time horizon used for emission metrics should nevertheless be consistent with that central goal of the Paris Agreement. We therefore support the use of the 20-year time horizon over the 100-year version, when binary choices between these two must be made, due to the better alignment of the former with the temperature goals of the Paris Agreement. However, to best align emission metrics with the Paris Agreement, we recommend the use of the 24-year time horizon, using 2045 as the endpoint time, with its associated $GWP_{1.5\,°C} = 75$ and $GTP_{1.5\,°C} = 41$.

**Methods**

Raw scenario data were taken from the IPCC SR1.5 Scenario Database[44]. These scenarios are "model-based climate change mitigation pathways" used in the IPCC's Special Report on Global Warming of 1.5 °C[51]. Scenarios were categorized by IAM, aggregating those with different versions of the same IAM (such as IMAGE 3.0.1 and IMAGE 3.0.2 being denoted simply 'IMAGE'). Ten IAMs were analyzed: GCAM, IMAGE, MERGE, MESSAGE, REMIND, WITCH, POLES, IEA, C-ROADS, and AIMS. The scenarios analyzed across these IAMs span a wide range of potential future pathways. These include emphases on bioenergy, sustainable development, negative emissions, carbon dioxide removal, and transport. While it is true that not all possible scenarios are compiled here, this set of scenarios forms the basis of the international climate community's current best understanding of realistic potential pathways. Median temperatures were taken from the variable "AR5 climate diagnostics|Temperature|Global Mean|MAGICC6|MED".

Emission metrics were calculated using data from the IPCC's Sixth Assessment Report[1,45]. These metrics use the most recent carbon-climate feedback impulse response functions[52] and radiative efficiencies[53]. Details on the calculation of GWP, GTP, iGTP, and CGTP are given below, based on the formulation in Gasser et al.[52]. In our work, we use the exact dataset used in AR6 (see Data Availability) for consistency with their Table 7.15. We incorporate the carbon cycle feedback but make no corrections for the carbon dioxide effects of different methane sources. Equivalent results can be found for specific types of methane (fossil or biogenic) in the Jupyter notebook (see Code Availability).

The GWP for a non-carbon dioxide greenhouse gas is the ratio of the absolute Global Warming Potential (AGWP$_X$) of gas X divided by the AGWP of carbon dioxide. For most greenhouse gases, including methane, the AGWP$_X$ at a time horizon $t$ is given by

$$AGWP_X(t) = \varphi^X \int_o^t r_Q^X(t)dt, \qquad (2)$$

where $\varphi^X$ is the radiative efficiency of gas X, and $r_Q^X(t)$ is the climate impulse response function, values and equations for which can be found in Appendices C and D of Gasser et al.[52]. The GWP for gas X (GWP$_X$) is then given by AGWP$_X$ / AGWP$_{CO2}$.

The GTP for a non-carbon dioxide greenhouse gas is calculated as the ratio of the absolute Global Temperature change Potential (AGTP$_X$) of gas X divided by the AGTP of carbon dioxide. For most greenhouse gases, the AGTP$_X$ at time horizon $t$ is given by

$$AGTP_X(t) = \varphi^X \lambda \int_o^t r_Q^X(t')r_T(t-t')dt', \qquad (3)$$

where $\lambda r_T(t-t')$, the climate response function, is given analytically in Appendix C of Gasser et al.[52]. The GTP$_X$ is then given by AGTP$_X$ / AGTP$_{CO2}$.

The integrated Global Temperature Change potential (iGTP$_X$) is the ratio found by integrating the AGTP$_X$ of a non-carbon dioxide gas X divided by the integrated AGTP of carbon dioxide[37]:

$$iGTP_X(t) = iAGTP_X(t) / iAGTP_{CO_2}(t)$$

$$= \int_0^t AGTP_X(t') dt' / \int_0^t AGTP_{CO_2}(t') dt'. \qquad (4)$$

The CombinedGTP (CGTP)[40] is the ratio of the temperature impact caused by a sustained increase in emission rate of a non-carbon dioxide gas divided by the temperature impact of a pulse of carbon dioxide of mass equal to the annual change in emission rate. This equating of a step function in non-carbon dioxide greenhouse gas emission rate with a pulse of carbon dioxide makes CGTP a 'step-pulse' emission metric with units of years. As outlined by Collins et al.[40], the order of integration can be switched and CGTP can be expressed as

$$CGTP_X(t) = iAGTP_X(t) / AGTP_{CO_2}(t). \tag{5}$$

To derive the relationships shown in Figure 3b, we simply inserted Equation 1 (the relationship between time and peak temperature) into Equations 2-5. This means that for any temperature goal ($T_{peak}$), Equation 1 gives a corresponding time of peak ($t_{peak}$) and associated time horizon from present day, which can then be inserted into the equations for AGWP, AGTP, and iAGTP for methane and carbon dioxide and then divided to give GWP, GTP, iGTP, and CGTP.

To calculate uncertainties for the parameters in Equation 1 and prediction intervals throughout, we used the Python package statsmodels. Implementation details can be found in the Python code that we have made available online (see Code availability).

**Data availability**

The data that support the findings of this study are openly available on the IPCC SR1.5 Scenario Database download page[44] and the IPCC AR6 data repository[1,45]. Our code repository (below) also contains the raw data required to replicate our analysis. Processed data used in the analysis and generation of figures and tables is available directly from the corresponding author on request or can be generated by running the provided Jupyter Python notebook (see Code availability).

**Code availability**

The Jupyter Python notebook used for the analysis, figures, and tables presented in this work can be downloaded from Zenodo at https://doi.org/10.5281/zenodo.4609302.

**Acknowledgements**

S. A. was supported by the National Sciences and Engineering Research Council of Canada, a Stanford Woods Institute Goldman Graduate Fellowship, and the Stanford Data Science Scholars program. S. A. and R. B. J. acknowledge support from the Stanford Woods Institute for the Environment (SPO 164153 WTAQE), and the Gordon and Betty Moore Foundation (Grant GBMF5439, "Advancing Understanding of the Global Methane Cycle"; Stanford University).